\documentclass[aps,preprint,eqsecnum,showpacs]{revtex4}
\usepackage{amsfonts}
\usepackage{amsmath}
\usepackage[latin1]{inputenc}
\usepackage[T1]{fontenc}
\usepackage{textcomp}
\usepackage{times}
\begin{document}
\title{Generalized Mittag--Leffler functions in the theory of
finite--size scaling for systems with strong anisotropy and/or
long--range interaction}
\author{H. Chamati}
\email{chamati@issp.bas.bg}
\author{N.S. Tonchev}
\email{tonchev@issp.bas.bg}
\affiliation{Institute of Solid State Physics, 72 Tzarigradsko Chauss\'ee,
1784 Sofia, Bulgaria}
\begin{abstract}
The difficulties arising in the investigation of finite--size scaling
in $d$--dimensional $O(n)$ systems with strong anisotropy and/or
long--range interaction, decaying with the interparticle distance $r$
as $r^{-d-\sigma}$ ($0<\sigma\leq2$), are discussed. Some integral
representations aiming at the simplification of the investigations
are presented for the classical and quantum lattice sums that take
place in the theory. Special attention is paid to a more general
form allowing to treat both cases on an equal footing and in
addition cases with strong anisotropic interactions and different
geometries. The analysis is simplified further by expressing this
general form in terms of a generalization of the Mittag--Leffler
special functions. This turned out to be very useful for the
extraction of asymptotic finite--size behaviours of the thermodynamic
functions.
\end{abstract}
\maketitle

\section{Introduction}
The \textit{standard} finite--size scaling theory (FSS) is
usually formulated in terms of {\it only one} reference length --
the bulk correlation length $\xi$. For a system with finite linear
size $L$, the main statements of the theory are:

{\bf (i)} The only relevant variable in terms of which the
properties of the finite system depend in the neighborhood of the
bulk critical parameter (the temperature in classical systems and
the corresponding quantum parameter in quantum systems) driving the
phase transition is $L/\xi$.

{\bf (ii)} The rounding of the thermodynamic function
exhibiting singularities at the bulk phase transition in a given
finite system sets in when $L/\xi=O(1)$.

The tacit assumption is that all other reference lengths  are
irrelevant and will lead {\it only to corrections} towards the above
picture. Moreover, the crucial point in the finite--size theory
is that we always assume: the {\it finite--size} linear $L$ of
systems under consideration and the correlation length $\xi$ are
large in the microscopic scale. This means $ L\gg a$ and  $\xi \gg
a$, where $a$ is the lattice spacing. For a comprehensive recent
review on this subject and other studies related to it see Ref.
\cite{brankov2000}

To investigate the finite--size scaling properties of systems with
long--range (LR) interaction decaying at large distances $r$ as
$r^{-d-\sigma}$, where $d$ is the space dimensionality and
$0<\sigma\leq2$ a parameter controlling the range of the
interaction, one needs to extract the finite--size effects from the
$d$ dimensional lattice sum
\begin{equation}\label{fsssum}
W_{d,\sigma}^\gamma(t,L)=\sum_{\mathbf q} \frac1{\left(t+|{\mathbf
q}|^\sigma\right)^\gamma},
\end{equation}
where due to the periodic boundary conditions $\mathbf q$ is a
discrete vector with components $q_i=2\pi n_i/L$ ($n_i=0,\pm1,\pm2,
\cdots)$, $i=1,\cdots,d$. $L$ is the size of the box confining the
system and $t$ is a parameter measuring the distance to the bulk
critical point. Here we will not comment on the
restrictions on $d$, $\sigma$ and $\gamma$, nor will we dwell on the
problem of convergence of (\ref{fsssum}); these details should be
clear from the context. Notice that $\sigma=2$ corresponds formally
to the case of short--range (SR) interaction \cite{GZ}. $\gamma$
is a parameter allowing to treat classical $(\gamma=1)$ and quantum
$(\gamma=1/2)$ systems on an equal footing. Higher values of
$\gamma$ appear in the investigation of finite--size systems to the
one loop order in the field theoretical approach. For a recent
review on the critical properties of systems with LR
interaction see Ref. \cite{chamati2003,tonchev2004}.

An other reason for considering $W_{d,\sigma}^\gamma(t,L)$ is, as we
will see below, the direct mapping between the lattice sum
(\ref{fsssum}) and some lattice sums in combinations with integrals
that appear in the theory of systems with strong anisotropic LR
interaction of the asymptotic form
(see for example references \cite{BW89,chamati2000,HS01,H02,CGGP})
\begin{equation}\label{ai}
J({\bf q})\simeq J(0) +a_{||}|{\bf
q}_{||}|^{\rho}+a_{\perp}|{\bf q}_{\perp}|^{\sigma},
\end{equation}
where the first $r$ directions (called ``parallel'' and denoted by
the subscript $||$) are extended to infinity and the remaining $s$
directions (called ``transverse'' and denoted by $\perp$) are kept
finite, with $r+s=d$ and $a_{\perp}$, $a_{||}$ are metric factors
and $\rho,\sigma>0$.
Let us note that there is a limited number of papers
that consider FSS assumptions on a microscopic models
\cite{chamati2000,CGGP}. It seems that
the considerations are mainly on phenomenological level or via
computer simulations (see, for example refs. \cite{BW89,HS01,H02})
because of the problems emerging  in analytical treatments.

The study of the difference between the $d$--dimensional sum
(\ref{fsssum}) at large sizes $L$ and its limiting integrals is
crucial in the derivation of finite--size effects. In the particular
cases $\gamma=1$ or $1/2$ to solve this problem several approaches
have been proposed
\cite{fisher1986,brankov88,BD91,brankov89,singh89,privman90,
chamati1994,chamati2000,luijten99}.
Among them the most {\it universal} one is that
based on the Poisson summation formula
\cite{brankov88,BD91,brankov89,chamati1994,chamati2000,chamati2000b}.
The aim of
this approach \cite{B} is to factorize  the $d$--dimensional sum in
the r.h.s of equation~(\ref{fsssum}) and to reduce it to an
one--dimensional effective problem. The term $|\mathbf{q}|^\sigma$ in
conjunction with $\gamma$ to be arbitrary in the interval $0<\gamma
<\infty $ causes peculiar mathematical problems concerning the
evaluation of the lattice sums over $\mathbf{q}$. The aim of the
present study is to generalize the previous investigations for
arbitrary $0<\gamma <\infty $. By virtue of the relation
\begin{equation}
W_{d,\sigma}^{\gamma+n}(t,L)=\frac{(-1)^{n}}{\gamma(\gamma
+1)\cdots(\gamma +n-1)}\frac{d^{n}}{d^{n}t}W_{d,\sigma}^\gamma(t,L),
\end{equation}
one can see that it is formally sufficient to consider only the case
$0<\gamma \leq 1$. Let us first consider separately
the classical and the quantum case.

\subsection{Classical case ($\gamma=1$)}\label{classicalcase}
In the case of classical systems with SR interaction,
corresponding to $\gamma=1$ and $\sigma=2$, the following
substitution is used as an indispensable ingredient for the
FSS calculations (see, e.g. \cite{brankov2000}).
\begin{equation}\label{Scr}
W_{d,2}^1(t,L) =\int_{0}^{\infty}dx\exp(-tx)
\left[\sum_{q}\exp(-q^{2}x)\right]^{d},
\end{equation}
where $q$ is one--dimensional discrete vector.

This is the so called Schwinger parametric representation. The analytic
properties of the function $\sum_{q}\exp(-q^{2}x)$ are very well known,
since it is nothing but the reduced Jacobi $\theta_{3}$ function. The
aim of the above procedure is two--fold: (i) to exponentiate the summand
and to reduce the $d$--dimensional sum to a one--dimensional sum with
well known analytic properties, and (ii) to give the dimensionality $d$
the status of a continuous variable.

In the presence of $\mathbf{q}^{\sigma}$ term, it is not so easy to
realize this procedure. The problem has been solved by suggesting
different generalizations \cite{CGGP,brankov88,privman90,luijten99,BD91}
of the Schwinger representation (\ref{Scr}) that lead to different
obstacles.

In order to preserve the possibility for further analytical
consideration  based on the properties of the reduced Jacobi
$\theta_{3}$ function in Ref. \cite{brankov88} the following
representation has been used
\begin{equation}\label{BTr}
W_{d,\sigma}^1(t,L)= t^{\frac{2-\sigma}{\sigma}} \int_{0}^{\infty}dx
Q_{\sigma}(t^{2/\sigma}x)\left[\sum_{q}\exp(-q^{2}x)\right]^{d}.
\end{equation}
The price one pays for this is that instead of the simple
exponent in the integrand of (\ref{Scr}), the function
$Q_{\sigma}(t)$ appears:
\begin{subequations}\label{F}
\begin{equation}
Q_{\sigma}(x)= \int_{0}^{\infty}dy\exp(-xy){\tilde
Q}_{\sigma}(y),
\end{equation}
where
\begin{equation}
{\tilde Q}_{\sigma}(y)=\frac{1}{\pi}
\frac{\sin\left(\frac\sigma2\pi\right)y^{\frac\sigma2}}
{1+2y^{\frac\sigma2}\cos\left(\frac\sigma2\pi\right)+y^{\sigma}},
\quad 0<\sigma<2.
\end{equation}
\end{subequations}
First the connection between $Q_\sigma(x)$ and the Mittag--Leffler type
functions in the theory of FSS was established in reference
\cite{brankov89}. This reads
\begin{equation}\label{MLFa}
Q_{\sigma}(x)=x^{\frac\sigma2 -
1}E_{\frac\sigma2,\frac\sigma2}(-x^{\frac\sigma2}),
\end{equation}
where $E_{\alpha,\beta}(z)$ are entire functions of the Mittag--Leffler
type defined by the power series \cite{bateman1955,dzherbashyan1993}
\begin{equation}\label{defMLa}
E_{\alpha,\beta}(z)=\sum_{k=0}^{\infty}\frac{z^{k}}{\Gamma(\alpha
k+\beta)}, \ \ \ \ \ \alpha, \beta \in \mathbb{C}, \ \ \ \ {\mathrm
Re}(\alpha)>0.
\end{equation}
This shows that the study of the finite--size behaviour of the lattice
sum
\begin{equation}\label{brankov}
W_{d,\sigma}^1(t,L)=
\int_{0}^{\infty}dx x^{\frac\sigma2 -
1}E_{\frac\sigma2,\frac\sigma2}(-tx^{\frac\sigma2})
\left[\sum_{q}\exp(-q^{2}x)\right]^{d},
\end{equation}
are a direct consequence of the analytical properties of
$E_{\alpha,\beta}(z)$ \cite{brankov89}.

\subsection{Pure quantum case $(\gamma=1/2)$}\label{quantumcase}
For the investigation of the FSS at zero temperature when the phase
transition is driven by a quantum parameter, in (\ref{fsssum}) we
have $\gamma=\frac12$.
Then the following integral representation is obtained
\begin{equation}\label{qScr1}
W_{d,\sigma}^{\frac12}(t,L)=
\frac{2}{\pi}\int_{0}^{\infty}dp
\sum_{\mathbf{q}}\frac{1}{t+p^{2}+|\mathbf{q}|^{\sigma}}.
\end{equation}
The auxiliary variable $p^{2}$ adds an effective extra dimension.
Indeed the pure quantum system corresponds to a $d+1$ dimensional
anisotropic classical system with the geometry of a cylinder
$L^{d}\times\infty$. Recall that the bulk critical behaviour (e.g.
critical exponents) of a pure quantum system is equivalent to a
$d+z$ ($z=\frac\sigma2$ dynamic critical exponent) dimensional
classical system (this is the so called classical to quantum
crossover).

On the other hand, in the spirit of  relation (\ref{brankov})
the following modification  has been proposed \cite{chamati1994}
\begin{equation}\label{BTrH}
W_{d,\sigma}^{\frac12}(t,L)= \int_{0}^{\infty}dx x^{\frac\sigma4-1}
G_{\frac\sigma2,\frac\sigma4}(-tx^{\frac\sigma2})
\left[\sum_{q}\exp(-q^{2}x)\right]^{d}.
\end{equation}
The new functions $G_{\alpha,\beta}(z)$ are defined by the
power series \cite{chamati1994}
\begin{equation}\label{defML}
G_{\alpha,\beta}(z)=\frac{1}{\sqrt{\pi}}\sum_{k=0}^{\infty}
\frac{\Gamma(k+\frac12)}{\Gamma(\alpha k+\beta)}\frac{z^k}{k!}, \ \
\ \ \ \alpha, \beta \in \mathbb{C}, \ \ \ \ {\mathrm Re}(\alpha)>0.
\end{equation}
Some results on the analytic behaviour of these functions are
presented in reference \cite{chamati2000}. In the particular case
$\alpha=\frac\sigma2$, $\beta=\frac\sigma4$ the following identity
\cite{tonchev2004}
\begin{equation}\label{NT}
G_{\frac\sigma2,\frac\sigma4}(-z)=\frac{2}{\pi}\int_{0}^{\infty}
E_{\frac\sigma2,\frac\sigma2}\left(-(z+p^{2})\right)dp,
\end{equation}
can be obtained from Eq. (\ref{qScr1}) and the relation of its
l.h.s and r.h.s. with the functions $G_{\frac\sigma2,\frac\sigma4}(z)$
and $E_{\frac\sigma2,\frac\sigma2}(z)$, respectively.

\subsection{Anisotropic case $(0<\gamma<1)$}
In this case, instead of (\ref{qScr1}) we propose the following
identity that can be obtained (see Appendix \ref{appendix}) after
some algebra $(0<\gamma<1)$
\begin{equation}\label{qScr1a}
W_{d,\sigma}^\gamma(t,L)=
\frac{1}{(1-\gamma)\Gamma(\gamma)\Gamma(1-\gamma)}
\int_{0}^{\infty}dp\sum_{\mathbf{q}}\frac{1}{t
+p^{\frac{1}{1-\gamma}}+|\mathbf{q}|^{\sigma}}.
\end{equation}

Equation (\ref{qScr1a}) generalizes the result (\ref{qScr1})
corresponding to the pure quantum case. Here the auxiliary variable
$p^{\frac{1}{1-\gamma}}$ acts effectively as an anisotropic extra
dimension that generates additional mathematical
difficulties. In the denominator of the summand in
the r.h.s of Eq. (\ref{qScr1a}) one can easily recognize the form of the
anisotropic interaction (\ref{ai}) with $s=d, r=1$ and
$\rho=1/(1-\gamma)$.

In this paper, we present new representation formulas for the
lattice sums defined in equation (\ref{fsssum}) relevant to the
investigations of the finite--size scaling properties of a large class
of systems: classical, quantum and systems with strong anisotropy.
Following the lines of consideration mentioned
for the particular cases of the previous
subsections \ref{classicalcase} and \ref{quantumcase} our aim here is to
present functions depending on three parameters $\alpha,\beta$ and
$\gamma$ that can play the same role as the functions
$E_{\alpha,\beta}(z)$ and $G_{\alpha,\beta}(z)$. The mathematical
properties of these functions will be discussed in the next section
and some applications will be given.

\section{Generalized Mittag--Leffler functions}\label{app}
The following generalization of the Mittag--Leffler functions is
defined by the power series \cite{prabhakar1971}
\begin{equation}\label{gfunction}
E_{\alpha,\beta}^\gamma(z)=\sum_{k=0}^\infty\frac{(\gamma)_k}
{\Gamma(\alpha k+\beta)}\frac{z^k}{k!}, \ \ \ \ \
\alpha, \beta, \gamma \in \mathbb{C}, \ \ \ \ {\mathrm Re}(\alpha)>0.
\end{equation}
are of a significant interest. Here
\begin{equation}
(\gamma)_0=1,\ (\gamma)_k=\gamma(\gamma+1)(\gamma+2)\cdots(\gamma+k-1)
=\frac{\Gamma(k+\gamma)}{\Gamma(\gamma)}, \ \ \ \ k=1,2,\cdots.
\end{equation}
These functions are named after
Mittag--Leffler who first introduced the particular case with
$\beta=\gamma=1$. Recently the interest in this type of functions
has grown up by their applications in some evolution problems and
by their various generalizations appearing in the solution of
differential and integral equations. For some mathematical
applications see Refs. \cite{saxena2004,kilbas2004} and references
therein.

In the present study we will show that these functions can
play an intrinsic role in the theory of FSS. Remark that the
generalized functions (\ref{gfunction}) reduce to the
$G_{\alpha,\beta}$ given by (\ref{defML}) in the particular case
$\gamma=\frac12$.

One of the most striking properties of these functions is that they obey
the following identity
\begin{equation}\label{identity1}
(1+z)^{-\gamma}=\int_0^\infty dxe^{-x}x^{\beta-1}
E_{\alpha,\beta}^\gamma(-x^\alpha z),\qquad \mathrm{Re}(\gamma),
\mathrm{Re}(\beta)>0,\qquad |z|<1,
\end{equation}
which is obtained by means of term--by--term integration of the series
(\ref{gfunction}). As we will show the identity (\ref{identity1}) lies
in the basis of the mathematical investigation of FSS in systems with LR
interaction. If we set in the identity (\ref{identity1}) $z=y^{-\alpha},
y>0$, and $x=ty$, we will obtain the Laplace transform
\begin{equation}
\frac{y^{\alpha\gamma - \beta}}{(1+y^{\alpha})^{\gamma}}=
\int_0^\infty dte^{-yt}t^{\beta-1}
E_{\alpha,\beta}^\gamma(-t^\alpha )
\end{equation}
from which we derive a new identity by setting
$\beta=\alpha\gamma$:
\begin{equation}\label{identity}
\frac{1}{(1+y^{\alpha})^{\gamma}}= \int_0^\infty
dte^{-yt}t^{\alpha\gamma-1}
E_{\alpha,\alpha\gamma}^\gamma(-t^\alpha ).
\end{equation}
With the help of the above identity one immediately obtains the
relation
\begin{equation}\label{HCNT}
W_{d,\sigma}^\gamma(t,L)= \int_{0}^{\infty}dx x^{\gamma\frac\sigma2
-1} E^{\gamma}_{\frac\sigma2,\gamma\frac\sigma2}(-tx^{\frac\sigma2})
\left[\sum_{q}\exp(-q^{2}x)\right]^{d},\qquad \mathrm{Re}(\gamma)>0,
\end{equation}
which is the quested generalization of (\ref{brankov}) and
(\ref{BTrH}). Now it is easy to obtain from Eqs.
(\ref{brankov}), (\ref{qScr1a}) and (\ref{HCNT}) the generalization
of the integral relation (\ref{NT}):
\begin{equation}\label{HTG}
E^{\gamma}_{\frac\sigma2,\gamma\frac\sigma2}(-z)=
\frac{1}{(1-\gamma)\Gamma(\gamma)\Gamma(1-\gamma)} \int_{0}^{\infty}
E_{\frac\sigma2,\frac\sigma2}^{1}\left(-(z+p^{\frac{1}{1-\gamma}})\right)dp,
\qquad 0<\gamma<1.
\end{equation}
Notice that
$E_{\frac\sigma2,\frac\sigma2}^{1}:=E_{\frac\sigma2,\frac\sigma2}$.
The asymptotic expansion for $z\gg1$ of the generalized
Mittag--Leffler functions $E_{\alpha,\beta}^\gamma(z)$ can be
obtained from (see Appendix \ref{appendix1}):
\begin{equation}\label{infty}
E_{\alpha,\beta}^\gamma(-z)=\sum_{k=0}^\infty(-1)^k
\frac{(\gamma)_k}{\Gamma[\beta-\alpha(k+\gamma)]}
\frac{z^{-(k+\gamma)}}{k!}, \quad |z|>1.
\end{equation}

In the particular case $\beta=\alpha\gamma$, relevant to the physical
situations we are discussing here, equation (\ref{infty}) reduces to
\begin{equation}\label{infs}
E_{\alpha,\alpha\gamma}^\gamma(-x)\simeq
-\gamma\frac{x^{-(1+\gamma)}}{\Gamma(-\alpha)},\quad x\gg1,
\end{equation}
where only the leading term is accounted for.

In the following section we will discuss some applications and relations
to previous results obtained in the framework of the FSS investigations
in the classical and the quantum cases.

\section{Finite--Size computations}
For FSS computations of $O(n)$ systems one needs the large--$L$
behaviour of normalized lattice sums defined by
\begin{equation}\label{wpr}
\overline{W}_{d,\sigma}^\gamma(t,L)
=\frac1{L^d}\sum_{\mathbf{q}\neq\mathbf{0}}
\frac1{\left(t+|\mathbf{q}|^\sigma\right)^\gamma},
\end{equation}
where $\gamma$ is relevant to different physical cases. Indeed for
integer $\gamma$ we have classical systems, while for half--integers
we have the quantum situation and for $\gamma<1$ systems with
geometry $L^{r}\times\infty^{s}$ and strong anisotropy of the type
(\ref{ai}). The last follows directly from the identity (compare
with (\ref{qScr1a}))
\begin{equation}\label{qScr2a}
\int_{0}^{\infty}dp\sum_{\mathbf{q}}\frac{1}{t
+p^{\rho}+|\mathbf{q}|^{\sigma}}=\frac{\Gamma(1-\frac{1}
{\rho})\Gamma(\frac{1}{\rho})}{\rho}
W_{d,\sigma}^{1-\frac{1}{\rho}}(t,L),\qquad \rho>1.
\end{equation}

The method we use here to extract the large--$L$ behaviour of
(\ref{wpr}) is based upon the identity (\ref{HCNT}). After
rearrangement of (\ref{wpr}) we obtain
\begin{subequations}
\begin{equation}\label{xxx}
\overline{W}_{d,\sigma}^\gamma(t,L)=
\frac{L^{\gamma\sigma-d}}{(2\pi)^{\gamma\sigma}}
\int_0^\infty dxx^{\gamma\frac\sigma2-1}
E^\gamma_{\frac\sigma2,\gamma\frac\sigma2}
\left(-\frac{tL^\sigma}{(2\pi)^\sigma}x^\frac\sigma2\right)
\left[A^d(x)-1\right],
\end{equation}
where
\begin{equation}
A(x)\equiv\sum_{n=-\infty}^{+\infty}e^{-xn^2}.
\end{equation}
\end{subequations}
 For large $x$, $A(x)-1$ decreases exponentially and the
integral in the right--hand side of equation (\ref{xxx}) converges at
infinity. For $x\to0$, the Poisson transformation formula
\begin{equation}\label{poiss}
A(x)=\sqrt{\frac{\pi}x}A\left(\frac{\pi^2}x\right)
\end{equation}
shows that $A(x)$ converges.

For small $x$ the integral in the right--hand side of equation
(\ref{xxx}) has an ultraviolet divergence for
$\mathrm{Re}(d)>\gamma\sigma$. So, an analytic continuation in $d$
is required to give a meaning to the integral. Adding and
subtracting the small behaviour of the function $A(x)$, we get after
straightforward algebra,
\begin{subequations}
\begin{equation}\label{fsst}
\overline{W}_{d,\sigma}^\gamma(t,L)=
L^{-d+\gamma\sigma}\left[D_{d,\sigma}^\gamma\left(tL^\sigma\right)^{\frac{d-\gamma\sigma}
{\sigma}}+ F^\gamma_{d,\sigma}\left(tL^{\sigma}\right) \right]
\end{equation}
where the constant
\begin{equation}
D^\gamma_{d,\sigma}=\frac2\sigma\frac1{(4\pi)^{\frac d2}}
\frac{\Gamma\left(\gamma-\frac d\sigma\right) \Gamma\left(\frac
d\sigma\right)} {\Gamma(\gamma)\Gamma\left(\frac d2\right)}
\end{equation}
and the functions
\begin{equation}\label{fssshift}
F_{d,\sigma}^\gamma(y)= \frac1{(2\pi)^{\gamma\sigma}} \int_0^\infty
dxx^{\gamma\frac\sigma2-1}
E_{\frac\sigma2,\gamma\frac\sigma2}^\gamma\left(-\frac
y{(2\pi)^\sigma} x^{\frac\sigma2}\right)
\left[A^d(x)-1-\left(\frac\pi x\right)^{\frac d2}\right].
\end{equation}
\end{subequations}
The first term in (\ref{fsst}) is the bulk contribution (it is
$L$--independent) and the second term is the corresponding
finite--size correction. The form (\ref{fsst}) is suitable for
the investigation of FSS in the vicinity of the critical point i.e.
$t\simeq0$. The function $F_{d,\sigma}^\gamma(y)$ enters in the
expressions for the scaling functions of various thermodynamic
observabales. The dependence on the linear size $L$ of the system is
included in the scaling variable $y$. The behaviour of any
thermodynamic function is tightly related to the asymptotic behaviour of
$F_{d,\sigma}^\gamma(y)$, which in turn depend upon that of the
Mittag--Leffler
functions. For detailed discussions of
different models with the particular values $\gamma=1$ and
$\gamma=\frac12$, the reader is
invited to consult references
\cite{brankov2000,chamati2003,chamati1994,chamati2000,chamati2000b},
where the finite--size scaling predictions are investigated in great
details.
Let us not that at this level, the anisotropy of the scaling
behaviour in Eq. (\ref{fsst}) only appears through the
parameter $\gamma$.

By setting $t=0$ in (\ref{fsst}) we obtain an expression for the
finite--size shift of the bulk critical parameter driving the phase
transition. This is proportional to
$F_{d,\sigma}^\gamma(0)L^{-\lambda}$, where $\lambda=d-\gamma\sigma$
is the shift critical exponent for the specific value of $\gamma$.
The coefficient $F_{d,\sigma}^\gamma(0)$ can be evaluated
analytically as well as numerically for different values of the free
parameters $d$, $\sigma$ and $\gamma$ using the method developed in
reference \cite{chamati2000l}.

According to the standard FSS we {\it must} have
$\lambda=1/\nu$, where $1/\nu$ is the critical exponent measuring
the divergence of the correlation length. The value of $\nu$ depends
on the concrete microscopic model. For illustration, in the
particular case of symmetric $O(n)$ model in the limit $n\to
\infty$, one can consider two cases: classical and quantum, where
$\nu=1/(d-\sigma)$ and $\nu=1/(d-\sigma/2)$, respectively. In both
cases our result confirms the FSS theory predictions (see, e.g.
\cite{brankov2000}). Furthermore in order to make contact with the
case of strong anisotropy of refs. \cite{CGGP,T05} the effective
dimensionality $D=2d/\sigma+2/\rho$ must be introduced. It
determines the conditions for the phase transition to take place, if
the anisotropic LR  has the asymptotic form (\ref{ai}) with $s=1,
r=d$. In this case $\nu=2/(\sigma(D-2))$ and again we have agreement
with FSS theory , provided $2<D<4$. Moreover, introducing the
effective  dimension $D(\gamma)=2d/ \sigma+2(1-\gamma)$ we can
consider a more general classical system that includes $\gamma=1,
1/2, (1-1/\rho)$ as particular cases.

The finite--size correction to the bulk critical behaviour ( of e.g.
susceptibility)  can be extracted from the asymptotic form of the
functions $F_{d,\sigma}^\gamma(y)$ defined by (\ref{fssshift}) at
large argument. This in turn can be obtained with the help of the
expansion (\ref{infs}). After some algebra we get (see Appendix
\ref{appc})
\begin{subequations}\label{assym}
\begin{equation}
F_{d,2}^\gamma(y)\simeq-y^{-\gamma} +\left[\frac
d{2^{\gamma}(2\pi)^{\frac{d-1}2}\Gamma(\gamma)}\right]
y^{\frac14(d-2\gamma-1)}e^{-\sqrt{y}}
\end{equation}
for $\sigma=2$ i.e. for the SR case and
\begin{equation}
F_{d,\sigma}^\gamma(y)\simeq-y^{-\gamma}
+\left[2^\sigma\gamma\pi^{-\frac d2}
\frac{\Gamma\left(\frac{d+\sigma}2\right)}
{\Gamma\left(-\frac\sigma2\right)}
\sum_{\mathbf{l}\neq\mathbf{0}}\frac1{|\mathbf{l}|^{d+\sigma}}\right]y^{-(1+\gamma)}
\end{equation}
\end{subequations}
for $0<\sigma<2$, corresponding to the LR case. Equations
(\ref{assym}) generalize equations (3.29) of
reference \cite{chamati2000b} obtained for the particular case $\gamma=1$.

Equations (\ref{assym}) reflect the fact that for systems with LR
interaction the finite--size corrections fall--off in {\it power
law} rather than exponential as it is the case for their
counterparts with SR interaction. These results are generalizations
of those obtained previously in the case of classical systems
\cite{brankov89,BD91,singh89,chamati2000b} and those obtained for
their quantum counterparts \cite{chamati1994,chamati2000}. The
former cases can be obtained by using integer values for $\gamma$
and the latter ones by using  half--integer values.

\section{conclusion}
We presented some mathematical results on the investigation of the
FSS in $O(n)$ systems based on the generalized Mittag--Leffler
functions (\ref{gfunction}) that have well known analytic
properties. Mainly two type of systems are of particular interest.

\textbf{(i)} The fully finite $d$--dimensional systems with LR
interaction decaying algebraically with the interparticle distance.

This is the case with $0<\sigma<2$ in (\ref{fsssum}). A special
emphasis on the mathematical difficulties arising in the
investigation of the FSS both in the classical (with $\gamma=1$
in (\ref{fsssum})) and quantum cases (with $\gamma=1/2$ in
(\ref{fsssum})) are discussed. The used techniques allow the
investigations to be simplified and express the results for various
thermodynamic quantities in terms of simple, with well defined
analytic properties, mathematical functions. An integral
representation (\ref{HCNT}) to deal with such difficulties, at least
asymptotically, are presented. It turned out that both cases can be
treated on an equal footing.

\textbf{(ii)} The classical system with mixed geometries with both
finite and infinite sizes and strongly anisotropic interaction of
the type (\ref{ai}).

Such type of systems are considered in reference \cite{CGGP}, where
$0<\rho,\sigma<2$ and $\gamma=1-1/\rho$. An other interesting case
is the $m$--fold Lifshitz point that is characterized by an
instability associated with the absence of quadratic terms in the
form $q^{2}_{\alpha}$ in the effective Landau--Ginzburg--Wilson
Hamiltonian for all $\alpha=1,2,\cdots,n<d$\ \cite{BW89,cardy96}.
Then in (\ref{ai}) one must set $\rho=4,\sigma=2$ and $\gamma=3/4$.
This gives a simpler way of solving such problems using generalized
Mittag--Leffler functions. It is based on the established mapping
(\ref{qScr2a}) to a fully finite--size system with specific
$0<\gamma<1$ in (\ref{fsssum}).

In conclusion, our considerations establish that we
can study finite--size scaling behaviour of classical systems, quantum
systems and systems with strong anisotropy confined in mixed
geometry $\infty\times L^{d}$,  in the framework of a fully finite
anisotropic system with a {\it classical} critical behaviour. This is
achieved in a unified fashion, varying the superscript $\gamma$ in
the generalized Mittag--Leffler functions.

\begin{acknowledgments}
This work is supported by the Bulgarian Science Foundation under
Project F--1402.
\end{acknowledgments}

\appendix
\section{Derivation of equation (\ref{qScr1a})}\label{appendix}
We have the relation
\begin{equation}
\frac{1}{\mu}p^{-\alpha/\mu}\Gamma\left(\frac{\alpha}{\mu}\right)=
\int_{0}^{\infty}x^{\alpha-1} e^{-px^{\mu}}dx;\ \ \qquad \mu, \
\mathrm{Re}(\alpha),\ \mathrm{Re}(p)>0.
\end{equation}
or if $\alpha/\mu=\gamma$
\begin{equation}\label{g}
p^{-\gamma}=\frac{\mu}{\Gamma(\gamma)}\int_{0}^{\infty}x^{-(1-\gamma\mu)}
e^{-px^{\mu}}dx.
\end{equation}
Using twice (\ref{g}) we get
\begin{equation}
\label{gg}
\frac{1}{p^{\gamma}}=\frac{\mu}{\Gamma(\gamma)}
\frac{\nu}{\Gamma(1-\gamma\mu)}\int_{0}^{\infty}dx
\int_{0}^{\infty}dt
e^{-px^{\mu}}e^{-xt^{\nu}}t^{(1-\gamma\mu)\nu-1}.
\end{equation}
Now on the free parameters $\mu$ and $\nu$ we impose the
conditions
\begin{equation}
(1-\gamma\mu)\nu-1=0;\qquad \mu=1
\end{equation}
and obtain the identity ($\gamma<1$)
\begin{equation}
\frac{1}{p^{\gamma}}=\frac{1}{(1-\gamma)\Gamma(\gamma)\Gamma(1-\gamma)}
\int_{0}^{\infty}dt\frac{1}{p+t^{\frac{1}{1-\gamma}}}.
\end{equation}
Eq. (\ref{qScr1a}) immediately follows from the above identity.

\section{Derivation of the asymptotic behaviour of the generalized
Mittag--Leffler functions (\ref{infty})}\label{appendix1} An integral
representation of the generalized Mittag--Leffler functions
$E_{\alpha,\beta}^\gamma(z)$ can be obtained with the aid of the
Henkel integral for the inverse gamma function
\begin{equation}\label{hankel}
\frac1{\Gamma(z)}=\frac1{2\pi i}\int_{C}e^u u^{-z}dz,
\end{equation}
where the integration contour $C$ is a loop which starts and ends at
$x=-\infty$ and encircles the origin in the positive sense:
$-\pi\leq\mathrm{arg}\ z\leq\pi$ on $C$~\cite{abramovitz64}. This enables
to get the result
\begin{equation}\label{intrep}
E_{\alpha,\beta}^\gamma(z)=\frac1{2i\pi\alpha}\int_{C}dv
\frac{e^{v^{1/\alpha}}v^{\gamma-1+(1-\beta)/\alpha}}{(v-z)^{\gamma}}.
\end{equation}
In the following we will investigate the asymptotic behaviour of the
generalized Mittag--Leffler functions at large argument, following
the method used in reference \cite{chamati2000}. This may be
performed with the aid of the series \cite{PBM}
\begin{equation}\label{asse}
(x+z)^{-\gamma}=x^{-\gamma}\sum_{k=0}^{\infty}(-1)^k\frac{(\gamma)_k}{k!}
\left(\frac zx\right)^k,\qquad \mid\frac{z}{x}\mid\leq 1;\quad
\frac{z}{x}\neq -1.
\end{equation}
After substitution of the latter equation into the integral
representation (\ref{intrep}) one obtains (\ref{infty}).

\section{Large asymptotic behaviour of
$F_{\lowercase{d},\sigma}^\gamma(y)$ from (\ref{fssshift})}\label{appc}
To obtain the large $y$ asymptotic behaviour~(\ref{assym}) of the
functions
$F_{d,\sigma}^\gamma(y)$ we rewrite (\ref{fssshift}), with the help of
the identity (\ref{poiss}), in the form
\begin{eqnarray}\label{other}
F_{d,\sigma}^\gamma(y)&=&\frac{\pi^\frac d2}{(2\pi)^{\gamma\sigma}}
\int_0^\infty dx x^{\gamma\frac\sigma2-\frac d2-1}
E_{\frac\sigma2,\gamma\frac\sigma2}^\gamma
\left(-y\frac{x^\frac\sigma2}{(2\pi)^\sigma}\right)
\left[A^d\left(\frac{\pi^2}x\right)-1\right] \nonumber \\
&&-\frac1{(2\pi)^{\gamma\sigma}}\int_0^\infty dx x^{\gamma\frac\sigma2-1}
E^\gamma_{\frac\sigma2,\gamma\frac\sigma2}
\left(-y\frac{x^\frac\sigma2}{(2\pi)^\sigma}\right).
\end{eqnarray}
Using the identity
\begin{equation}\label{inti}
\int_0^\infty dx x^{\gamma\frac\sigma2-1}
E^\gamma_{\frac\sigma2,\gamma\frac\sigma2}
\left(-x^{\frac\sigma2}\right)=1, \ \ \ \  \sigma>0
\end{equation}
from the second term of equation (\ref{other}) we obtain the first
terms of equations (\ref{assym}).

Further, taking into account the asymptotic behaviour (\ref{infty}) of
the functions $E^\gamma_{\alpha,\beta}(z)$ and after subsequent integration
in the first term of equation (\ref{other}), we obtain finally the asymptotic
behaviour given by equations (\ref{assym}).

\end{document}